\documentstyle[12pt]{article}
\begin{document}

\def\lsim{\mathrel{\rlap{\lower3pt\hbox{\hskip0pt$\sim$}}
    \raise1pt\hbox{$<$}}}         
\def\gsim{\mathrel{\rlap{\lower4pt\hbox{\hskip1pt$\sim$}}
    \raise1pt\hbox{$>$}}}         
\def\dblint{\mathop{\rlap{\hbox{$\displaystyle\!\int\!\!\!\!\!\int$}
}
    \hbox{$\bigcirc$}}}
\def\ut#1{$\underline{\smash{\vphantom{y}\hbox{#1}}}$}

\newcommand{\beq}{\begin{equation}}
\newcommand{\eeq}{\end{equation}}
\newcommand{\aver}[1]{\langle #1\rangle}

\newcommand{\La}{\overline{\Lambda}}
\newcommand{\Lam}{\Lambda_{QCD}}

\newcommand{\lhs}{{\em lhs} }
\newcommand{\rhs}{{\em rhs} }

\newcommand{\ind}[1]{_{\begin{small}\mbox{#1}\end{small}}}
\newcommand{\hscale}{\mu\ind{hadr}}

\newcommand{\appa}{\mbox{\ae}}
\newcommand{\CP}{{\em CP } }
\newcommand{\fy}{\varphi}
\newcommand{\hi}{\chi}
\newcommand{\al}{\alpha}
\newcommand{\as}{\alpha_s}
\newcommand{\gf}{\gamma_5}
\newcommand{\gm}{\gamma_\mu}
\newcommand{\gn}{\gamma_\nu}
\newcommand{\be}{\beta}
\newcommand{\ga}{\gamma}
\newcommand{\de}{\delta}
\renewcommand{\Im}{\mbox{Im}\,}
\renewcommand{\Re}{\mbox{Re}\,}
\newcommand{\GeV}{\,\mbox{GeV}}
\newcommand{\MeV}{\,\mbox{MeV}}
\newcommand{\matel}[3]{\langle #1|#2|#3\rangle}
\newcommand{\state}[1]{|#1\rangle}
\newcommand{\ra}{\rightarrow}
\newcommand{\ve}[1]{\vec{\bf #1}}

\newcommand{\eq}[1]{eq.\hspace*{.1em}(\ref{#1}) }
\newcommand{\eqs}[1]{eqs.\hspace*{.1em}(\ref{#1}) }

\newcommand{\re}[1]{Ref.~\cite{#1}}
\newcommand{\res}[1]{Refs.~\cite{#1}}

\begin{titlepage}
\renewcommand{\thefootnote}{\fnsymbol{footnote}}

\begin{center} \Large
{\bf St.Petersburg Nuclear Physics Institute}\\
\end{center}
\begin{flushright}
PNPI-TH/b-9/94\\
hep-ph/yymmnnn\\
September 1994
\end{flushright}
\vspace{.3cm}
\begin{center}
\Large
{\bf Comment on the Renormalization Group Improvement\\
in Exclusive $b\ra c $ Transitions\\}
\end{center}
\vspace*{1cm}

\begin{center}
{\Large
N.G. Uraltsev} \\
\vspace{.4cm}
{\normalsize {\it St. Petersburg Nuclear Physics Institute,
Gatchina, St. Petersburg 188350, Russia}\\
\vspace*{3cm}}
{\Large{\bf Abstract}}
\end{center}
\vspace*{.2cm}

Using rather general consideration I argue that the numerical impact of
the existing next-to-leading logarithmic summations of terms $\log{m_b/m_c}$
for perturbative factors $\eta_A$, $\eta_V$ in the exclusive zero recoil
semileptonic transitions is irrelevant, and adopting corresponding
corrections beyond the exact one loop result for both estimating the values
of these formfactors and their theoretical uncertainty, is misleading.
The central theoretical value if taken literally is then $\eta_A\simeq 0.97$.

\end{titlepage}

1. Weak decays of beauty particles allow one the accurate theoretical
description based on QCD, which utilizes the existence of the large mass
scale $m_b$; this description in many instances is model-independent to a
large extent. The first step in a practical incarnation of this idea is to
account for the interaction physics associated with the high scale, which
can be done in a standard way expanding in $\as(m_Q)$. Typically the first
radiative corrections $\sim \as(m_b)$ are just those effects that produce
the largest impact in decays of $b$ hadrons, at least in the absence of
accidental cancellations.
Among the exclusive decays, most attention has been paid to $b\ra c $
transitions, and in particular in the zero recoil kinematics, where the most
informative conclusions can be done. For example, rather accurate
experimental determination of the KM mixing parameter $|V_{cb}|$ is possible
using the decay rate of $B\ra D^*\,\ell \nu$ (and, possibly,
$B\ra D\,\ell \nu$) extrapolated to the zero recoil point. The corresponding
theoretical accuracy of such determination can be as high as $10\%$ or even
somewhat better, following only the way to extract $|V_{cb}|$ from inclusive
semileptonic decay widths \cite{optsr,SUV}.

Perturbative radiative corrections to the $b\ra c$ semileptonic amplitude at
zero recoil have been calculated in one loop at least seven years ago
\cite{SV} and appeared to be rather small, at the level of $3\%\:$:
$$
\eta_A=1+\frac{\as}{\pi}\left(\frac{m_b+m_c}{m_b-m_c}\log{\frac{m_b}{m_c}}-
\frac{8}{3}\right)
$$
\beq
\eta_V=1+\frac{\as}{\pi}\left(\frac{m_b+m_c}{m_b-m_c}\log{\frac{m_b}{m_c}}-
2\right)
\label{1}
\eeq
where subscripts $V$ and $A$ mark vector and axial currents. In particular,
for $\eta_A$ this was a result of a partial cancellation that existed for
the real masses of $b$ and $c$ quarks. In $b\ra c$ transitions there exists
the region of loop momenta between $m_c$ and $m_b$ that contributes to the
renormalization of the matrix elements of weak currents, which happens to
yield the correction of the sign opposite to the correction to $\eta_A$ at
equal masses. It is obvious, though, that the numerical cancellation in the
first order correction does not mean that the whole perturbative one is also
suppressed; for there are next order corrections governed by the factor
$\as^2(m_c)$ that generically are as large as the literal one loop
correction.
Later some formal renormalization group improvement of the V-S calculation
has been done where next-to-leading corrections have been summed up in the
logarithmic approximation, i.e. accounting for the terms
$\as^{n+1}\log^n{\frac{m_b}{m_c}}$.

In this note I discuss in more detail arguments pointed out recently in
\res{optsr,SUV}, which show the numerical irrelevance of this renormgroup
improvement for obtaining the actual values of $\eta_A$ and $\eta_V$, and
examplify the statement made there that using NLA-``improved'' expressions
for estimating the theoretical {\em uncertainty} in the perturbative
renormalization factors is misleading. More consistent treatment outlined
here allows one to obtain the numerical result of the NLA (next-to-leading
approximation) accuracy without explicit calculations. This more complete
discussion seems to be worthwhile in view of
some confusion appearing in
literature (see, e.g. \res{N}). I confirm that further theoretical
improvement of the one loop result for $\eta_A$ can come only from a true two
loop calculation, or at least if the two loop correction is
exactly calculated at $m_c=m_b$. \vspace*{.25cm}

2. The main object of our discussion will be two renormalization factors
$\eta_A$, $\eta_V$ defined perturbatively through the ``forward'' matrix
elements as follows:
$$
\matel{c(p')}{\bar c \gamma_\mu\gamma_5 b}{b(p)}=\eta_A\,
\bar c(p') \gamma_\mu\gamma_5 b(p)
$$
$$
\matel{c(p')}{\bar c \gamma_\mu b}{b(p)}=\eta_V\,
\bar c(p') \gamma_\mu b(p)
$$
\beq
p'/m_c=p/m_b
\label{2}
\eeq
where $c(p')$, $b(p)$ are the perturbative on shell heavy quark states (in
the {\it lhs}) and the corresponding spinors (in the {\it rhs}). It is
important for what follows that $\eta_{A,V}$ are infrared finite: this is a
consequence of the fact that inelastic decay probability vanishes at the
maximal recoil $q_{lept}^2=(m_b-m_c)^2$.

Straightforward calculation of the one loop quark diagrams leads to the
first order result \cite{SV}
$$
\eta_A=1+\frac{\as}{\pi}c^{(1)}_A\simeq 0.965\;\;\;\;\;\;
c^{(1)}_A=\frac{m_b+m_c}{m_b-m_c}\log{\frac{m_b}{m_c}}-
\frac{8}{3}
$$
\beq
\eta_V=1+\frac{\as}{\pi}c^{(1)}_V\simeq 1.02\;\;\;\;\;\;
c^{(1)}_V=\frac{m_b+m_c}{m_b-m_c}\log{\frac{m_b}{m_c}}-2
\label{3}
\eeq
The value of $\as$ in the one loop calculation must be avaluated at a scale
of the order of $m_c$ or $m_b$. The common point of view is that to
eliminate this scale ambiguity and improve the accuracy of the estimate one
has to make the summation of the terms $(\log{m_b/m_c})^n$ in higher orders
of perturbation theory. In real world the value of $\log{m_b/m_c}\simeq 1.2$
is not a large parameter. It is true, however, that the actual relevance of
such an expansion parameter is difficult to judge {\em a priori}. Before
turning to concrete calculations let us analyse the pertrubative series
appearing in the logarithmic summation, from a more general perspective.

The most simple observation which, as is shown below, underlies all numerics
in the problem, is that $\eta_{A,V}$  are to be symmetric as functions of
$m_b$ and $m_c$. Therefore it is advantageous to introduce the geometric
average mass
$$
\bar{m}=(m_cm_b)^{1/2}\;\;,
$$
\beq
m_b=x\bar{m}\;\;,\;\;m_c=\frac{1}{x}\bar{m}\;\;,\;\;x=\sqrt{m_b/m_c}
\label{4}
\eeq
and use $\bar{\al}_s\equiv \as(\bar{m})$ as an expansion parameter for
perturbative series. For the sake of definiteness I will discuss the case of
axial current  and omit subscript ``$A$'' in what follows, where it is
unimportant. The same consideration obviously applies to both cases.

Starting from the point $\bar{m}$ one needs to sum up powers of
$\log{m_b/\bar{m}}$ and $\log{\bar{m}/m_c}$ which both equal to $\log{x}$.
Using the above notations one then can write the generic expansion for
$\eta$ as
$$\eta=1+\frac{\as(\bar m)}{\pi}c^{(1)}(x) + f_0(\frac{\as(\bar
m)}{\pi}\log{x})+ \frac{\as(\bar m)}{\pi}f_1(\frac{\as(\bar
m)}{\pi}\log{x})+
$$
\beq +\left(\frac{\as(\bar m)}{\pi}\right)^2
f_2(\frac{\as(\bar m)}{\pi}\log{x}) +\ldots \;\;\;.
\label{5}
\eeq
$f_0$ is the
matter of the leading $\log$ approximation (LLA), $f_1$ is obtined in the
next-to-leading summation, etc. It is assumed in \eq{5} that $f_0(r)\sim
{\cal O}(r^2)$ and $f_1(r)\sim {\cal O}(r)$ at small $r$, for the
corresponding terms are explicitly accounted for by the exact zero and first
order coefficients.

The function $c^{(1)}(x)$ is, of course, symmetric under the transformation
$x\ra 1/x$. The fact that
\beq
\eta(\bar m,x)=\eta(\bar m, 1/x)
\label{6}
\eeq
therefore implies that all $f_i(x)$ are even functions of their arguments:
\beq
f_i(r)=f_i(-r)\;\;.
\label{7}
\eeq
If $f_i(r)$ were analytical at $r=0$, \eq{7} would have meant that all
``NLA''effects due to $f_1,$ $f_2$ \ldots produce practically vanishing
impact on $\eta$ at $\log{\sqrt{m_b/m_c}}\simeq 0.6$ and
$r=\frac{\bar{\al}_s}{\pi}\log{\sqrt{m_b/m_c}}\simeq 0.05$. For then the
expansion of $f_1(r)$ would have started with $r^2$ with the term
\beq
\sim \; \frac{\bar{\al}_s}{\pi}\cdot\left(\frac{\bar{\al}_s}
{\pi}\log{x}\right)^2\simeq
2\cdot 10^{-4}\;\;,
\label{8}
\eeq
which is
parametrically smaller than non-$\log$ $(\as/\pi)^2$ terms omitted in the
NLA, even irrespective of the exact value. However, the functions $f_i(r)$
obtained in a $\log$ summation {\em via} an expansion in $1/\log{(m_b/m_c)}$
are {\em not} analytical in general at $r=0$. It is examplified, in
particular, by the one loop expression: in terms of the variable $r$ its
mass dependent part is written as
\beq
\coth{\left(\frac{\pi}{\bar{\al}_s}r\right)}\cdot
2r
\label{coth}
\eeq which is nothing but $2|r|$ in the LLA approach.

In fact, the LLA function $f_0$ is well known. Exponentiating the one loop
expression \eq{3} one readily obtains the standard LLA asymptotics \cite{SV}
$$
\eta^{LLA}=\left(\frac{\as(m_c)}{\as(m_b)}\right)^{\frac{2}{b}}=
\left(\frac{1+b\frac{\bar\as}{2\pi}|\log{x}|}
{1-b\frac{\bar\as}{2\pi}|\log{x}|}\right)^{\frac{2}{b}}\;\;,
$$
\beq
b=\frac{11}{3}N_c-\frac{2}{3}n_f=\frac{25}{3}\;\;.
\label{9}
\eeq
Therefore,
\beq
f_0(r)=\left(\frac{1+\frac{b}{2}|r|}{1-\frac{b}{2}|r|}
\right)^{\frac{2}{b}}-1-2|r|\simeq
2r^2+\frac{697}{54}|r|^3+\frac{643}{27}r^4+\ldots
\label{10}
\eeq
where
it is explicitly shown that the term linear in $\as\log$ is absent (this is
a feature of the improved version of LLA used in \eq{5}, which includes the
first loop correction exactly). Numerically the leading term in the {\em
rhs} constitutes approximately $4.6\cdot 10^{-3}$ whereas the second one is
$1.4\cdot 10^{-3}$; though, obviously, both can be neglected for practical
purposes.

Now let us consider the NLA terms in \eq{5} represented by $f_1$. The
second, analytical term in the expansion of $f_1$ at small $r$ is already of
the order of $(\bar{\al}_s/\pi)^3\,\log^2{\sqrt{m_b/m_c}}\approx 2\cdot
10^{-4}$ and definitely is to be neglected. Therefore the only real impact
could be identified with the first term linear in $|\log{\frac{m_b}{m_c}}|$
in the expansion of $f_1$. However this term is explicitly non-analytical as
a function of masses at $m_c=m_b$ and, therefore, cannot give any
resemblence to the true correction.

Let us dwell on this particular point. Neglecting terms
$(\as/\pi)^3$ and higher in \eq{5}, which are of the order of $10^{-3}$,
$\eta$ can be written in the form
\beq
\eta-1-\frac{\bar{\al}_s}{\pi}c^{(1)}(x)-
2 \left(\frac{\bar{\al}_s}{\pi}\log{x}\right)^2=
\left(\frac{\bar{\al}_s}{\pi}\right)^2 c_2+
f_1^{(1)}\left(\frac{\bar{\al}_s}{\pi}\right)^2 |\log{x}| + {\cal O}(\as^3)
\label{11}
\eeq
where $c_2$ denotes the value of the second order correction to $\eta$ at
$m_b=m_c$ and $f_1^{(1)}$ is the coefficient in the first term of the
expansion of the NLA function $f_1$ at small argument. NLA pretends to
evaluate the difference in the left hand side calculating the leading
correction to it.

As was shown above, the only numerical impact of the ``NLA improvement''
for $b\ra c$ decays at zero recoil is the second term in the {\em rhs} of
\eq{11}. However, the difference in the {\em lhs} is an analytical function
at $m_b=m_c$. Therefore, either $f_1^{(1)}$ must vanish or, if it does not,
it cannot be trusted as being overwhelmed by neglected terms. For example,
the derivative of $\eta$ with respect to $x$ is a continuous function in any
particular order in $\as(\bar m)$, as well as the whole formfactor. If a
term in the series is not analytical in some region of positive $m_c/m_b$,
it must be subleading.

In the complete expansion in $1/\log{x}$ \eq{5} such non-analytical terms can
easily emerge, for example, in the form $\sqrt{1+\log^2{x}}$, or in a
way similar to \eq{coth}, therefore non-vanishing $f_1^{(1)}$ does not mean
that an arithmetic mistake has been made. However, $\log$ expansions like
\eq{5} are asymptotic only. If it appears that the non-analytical terms like
$|\log{x}|$ are dominant, it clearly signals that one has gone too far
beyond the asymptotic region so that the difference between the true value
of the function and its truncated series is already larger than the last
accounted term. It is not surprising at all that this happens for $b\ra c$
transitions: the relevant ``large'' parameter of the expansion,
$\log{\sqrt{m_b/m_c}}$ , appears to be only $0.6\,$!

Let us return for the moment to the LLA. Its naive application would
produce a large numerical correction to $\eta_A$,
$\;\delta_\eta\simeq\frac{\al_s}{\pi}\log{\frac{m_b}{m_c}}\simeq +0.10$
practically irrespective of whether one uses the whole LLA function
$(\as(m_c)/\as(m_b))^{6/25}-1$, or only its first term
$\frac{2\bar\al_s}{\pi}|\log{x}|$ --
for its other terms are safely below $1\%$. According to
the general reasoning above in reality this first term is to be practically
absent, and one gets much better approximation $\eta_A\approx 1$ merely
neglecting it at all! The exact calculation of the one loop correction
clearly supports our general conclusion. This failure of the naive LLA has
been erroneously interpreted in papers \cite{N} as a necessity and relevance
of the NLA $\log$ corrections obtained in a standard renormgroup treatment.
In fact the true correction at a few percent accuracy is given by the exact
one loop expression \eq{3} and, of course, is quadratic in
$\log{\sqrt{m_b/m_c}}$ being thus rather small. One can formally ``improve''
the naive LLA result using the more consistent LLA expansion according to
\eq{5} which makes use of the exact one loop coefficient. Then at least the
leading term of the remaining LLA correction, \eq{10}, has a proper
analytical form, and problems start with only the next term which is
practically negligible. Though it goes without saying that the size of both
these terms lies below the ``noise level'' and they cannot be taken
seriously under any circumstances.

The very same consideration applies to the ``NLA improvement'' of \res{N}.
The result obtained there has the form
$$
\eta_A=\appa^2\left\{1+\frac{\as(m_b)-\as(m_c)}{\pi}Z_4-
\frac{8\as(m_c)}{3\pi}+ \right.
$$
$$
\left. +\frac{1}{x^2}\left(\frac{25}{54}-\frac{14}{27}\appa^{-3}+
\frac{1}{18}\appa^{-4}+\frac{8}{3}\log{\appa}\right)
+\frac{4\as(\mu)}{\pi}\frac{\log{x}}{x^2(x^2-1)}\right\}\;\;,
$$
\beq
\appa=\left(\frac{\as(m_c)}{\as(m_b)}\right)^{3/25}\;\;,\;\;\;\;
Z_4\simeq-1.5608
\label{12}
\eeq
and $\mu$ is some scale between $m_c$ and $m_b$.
This expression, formally valid, as usually, at
$\log{m_b/m_c}\gg1\,$, $\;\as\cdot\log{m_b/m_c}\sim 1$, allows one to
determine the NLA function $f_1$ in \eq{5}. In particular at small argument
it would be
\beq
f_1(r)=(-\frac{25}{3}Z_4+\frac{50}{3}s-\frac{2596}{225})\cdot |r|
+{\cal O}(r^2)\;\;\;,\;\;\;\mu\equiv m_c^sm_b^{1-s}
\label{13}
\eeq
and the ``wrong'' non-analytical term is the leading
one here for any intermediate scale $\mu$ at $0\le s\le 1$. For this reason
it is clear that such numerical predictions for the perturbative corections
to $\eta_A$ beyond the exact first loop result have no any relevance to the
real value of $\eta_A$.  The true result that can be seen from real two loop
calculations (when they are made) is to be quadratic in
$\log{\sqrt{m_b/m_c}}$ and, therefore, absolutely different numerically from
the one following from \eq{12} -- or both must be smaller than neglected
terms.

Similar to the case of the LLA, the problem with the leading term in the NLA
can be cured by an explicit calculation of the two loop correction,
which would eliminate
the spurious part proportional to $|\log{x}|$ from the NLA
function $f_1$. However, it is clear that the exact second loop calculation
would provide one with enough accuracy {\em per se} and any further
summation is not needed then.

It is worth to emphasize that this is {\em not} that has been done in
\res{N} where the NLA result is adjusted to fit exactly the {\bf one} loop
correction when expanded in $\as$. In fact, only the first three terms in
the curly brackets in \eq{12} can (and have been) directly obtained
by summing
the leading $\log$s; other terms are a rather arbitrary extrapolation made
to get the correct one loop result when $\as\log$ is small.

The theoretical analysis above elucidate the underlying problem faced in
the first two \res{N}, that makes it impossible a consistent derivation of
\eq{12} in its literal form. For to have any practical relevance the
calculation must include terms like $m_c/m_b$ that vanish in any
order of the expansion in $1/\log{(m_b/m_c)}$ being exponentially
suppressed in this parameter. On the other hand, it is the presence of these
terms that allows the analytical formfactor to get non-analytical
leading term. Therefore their inclusion is to be mandatory if one considers
the case when $\log{\sqrt{m_b/m_c}}$ is not large, -- which seems to be
impossible directly withing the orthodoxal renormgroup expansion used in
\res{N}.

It is interesting to note that, as it follows from the symmetry arguments,
the main impact is expected from the value of the second order correction at
the symmetric point $m_c=m_b$; the dependence on the ratio takes the
quadratic form\footnote{Note that introducing the variable $\log{(m_b/m_c)}$
instead of $(\frac{m_b}{m_c}-1)$ increases the radius of convergence of the
Taylor expansion of the first loop coefficient from $0<\frac{m_b}{m_c}\le 2$
to $e^{-2\pi}\le \frac{m_b}{m_c}\le e^{2\pi}\simeq 500$. Then one can, for
example, use only the first non-trivial term
$\frac{1}{6}\log^2{\frac{m_b}{m_c}}$ in the expansion of $c^{(1)}$ in
\eqs{3} to get its value at $m_c \ne m_b$
with more than enough accuracy. The second order
coefficient as it would come from \eq{12} is much more singular as will be
seen shortly, that does not seem to reflect reality.}
$\sim
\log^2{\sqrt{\frac{m_b}{m_c}}}\simeq 0.35$ and may well appear to be weak
numerically. However, all corrections to the vector formfactor vanish at
this point, and therefore the proper one loop result for $\eta_V$ is
expected to be more reliable. \vspace*{0.25cm}

3. Let us now discuss the anatomy of numbers in \eq{12} that lead to the
precise numerical prediction quoted in \res{N}. It is easily checked that the
literal expression in \eq{12} is approximated by its perturbative expansion
in $\as/\pi$ through the second order with a better than $1\%$ accuracy; for
simplicity we then neglect terms $\sim (\as/\pi)^3$. The first order term
exactly coinsides with $c_A^{(1)}$, and keeping only this one would give the
numerical one loop result
$$
\eta_A\simeq 0.965
$$ for the parameters adopted in \res{N}: $m_c/m_b=0.3$ and
$\Lambda^{(4)}_{\overline{\rm MS}}=0.25\GeV$.
We then consider the second order
coefficient that follows from \eq{12}. It takes the form
\beq
c^{(2)}(x)=\left(2+\frac{55x^2+300s-133}{9x^2(x^2-1)}\right)\,\log^2{x}+
\frac{1386x^{-2}-1875Z_4-3007}{225}\,\log{x}
\label{31}
\eeq
at $x\ge 1$; if $m_c>m_b\,$ $\;c^{(2)}(x)$ is to be calculated using \eq{31}
with the substitution $x\ra 1/x$
(the first term in the brackets is the ``improved LLA'' one).
The scale parameter $s$ is defined
in \eq{13}.  Again it is advantageous to use the scale variable
$y=-\log{x}$; the plot of $c^{(2)}(y)$ as given by \eq{31} is
shown in Fig.1 for $s=1$ ($\mu=m_c$) and $s=0$ ($\mu=m_b$). The central
value of $c^{(2)}$ at $m_c/m_b\simeq 0.3$ is approximately $2.7$ which
translates into the correction $+0.016$ to $\eta_A$. As it is seen
from Fig.1 this value of $c^{(2)}$ is largely determined by the linear term
$\propto|y|$.  However, just this linear term cannot be present in the
formfactor!  Moreover, if one varies $s$ from $0$ to $1$ $\;c^{(2)}$ varies
by $\pm 0.8$ which produces a $\pm 0.005$ variation in $\eta_A$, i.e.
practically the whole of $0.006$ estimated in \res{N} as the theoretical
uncertainty in $\eta_A$. Therefore, the major part of the theoretical
uncertainty was deduced from the term that would be absent at all in a
consistent calculation.

The central value for $\eta_A$ given by \eq{12} at $s=1/2$ is
\footnote{It is quoted in \res{N} as $0.986$ although such a value is
obtained only at extreme values of {\em all} parameters, viz. $m_c/m_b=0.25$
and the ``worst'' choice of $\mu=m_c$, whereas the literal dependence on the
strong coupling is not pronounced.}
$$
\bar{\eta}_A=0.976
$$ and the difference with the one loop value is mainly given by the shift
$0.016$ due to the second order term (the remaining difference is given by
the $(\as/\pi)^3$ terms in \eq{12}, which are also {\em ad hoc}; in
particular, they vary extremely sharp). The larger difference, $0.02$ occurs
when $\mu=m_c$ is assumed (the scale preferred in \res{N}), and is almost
completely due to the ``wrong'' term in the second order coefficient. It is
curious to note  that if one decides to adjust the scale $\mu$ in \eq{12} in
such a way as to get rid of the unphysical fracture in $c^{(2)}(y)$ at
$m_b=m_c$, viz. take $\mu$ somewhat above $m_b\;$ ($s\simeq -0.09$) then the
difference between the one loop and the NLA expression \eq{12} appears to be
only about $0.003$
\footnote{Actually even this smallish difference is almost completely
saturated by the ``improved LLA'' term.},
i.e. essentially smaller than the accuracy
pretended on even in \res{N}. This is not surprizing, of course, for the
value of $c^{(2)}$ at $m_c=m_b$, which only plays a role for a smooth
function, is set zero.  Thus one can see that any noticeable deviation of
the estimates inferred from the NLA espression (\ref{12}), from the
``educated one loop'' result does not have any justification.\vspace*{.1cm}

What is wrong, conceptually, in the estimate based on \eq{12}? The main flaw
is that it assumes $c^{(2)}$ to vanish\footnote{This happens in fact
for any order perturbative coefficient stemming from \eq{12}, except for
the first loop one.} at $m_b=m_c$ (it is worth reminding that $c^{(2)}$
depends on the renormalization scheme used to define the strong coupling).
It is clear that in reality the main part of the second order correction to
$\eta_A$ is to be given just by $c^{(2)}(m_b=m_c)$. For this point is a
regular one for the formfactor and it must be a smooth {\em even} function
of $y$, therefore its variation up to the real value of the mass ratio is
not expected to be large for $y^2\simeq 0.4$. No apparent reason is seen why
this value cannot be as large as $\pm 4\;$ -- if \eq{12}
gives, though rather arbitrarily, just values of a similar size;
this would shift the estimate for $\eta_A$ by $\pm 0.025$.
>From this
perspective the claim made in the last of \res{N} that the corrections in
$c^{(2)}$ that are not accounted for by \eq{12}, are not expected to exceed
unity, seems to be not self-consistent.

The second failure of \eq{12} is that at small $y$ it has a large linear
in $|y|$ term that dominates the value of the function -- but cannot be there
in reality. This fact comes from the {\em ad hoc} method of calculating the
``exponential'' terms $\frac{\log{x}}{x^k}$ within renormgroup approach,
suggested in the first two \res{N}. At least in the problem under
consideration it appears to be inconsistent with general symmetry properties
of the formfactor; in particular it leads to a rather sharp variation in the
perturbative coefficients at $|y|\lsim 0.6$, which can hardly be expected
from true functions.

The arguments given above make it clear that not only an attempt to determine
a more precize value of $\eta_{A,V}$ by a next-to-leading summation of
$\log{(m_b/m_c)}$, but even an estimate of a possible theoretical uncertainty
varying parameters entering this calculation (the way adopted in the
original papers \cite{N}), are completely misleading. Instead, one could
have used, in principle, for the latter purpose the one loop result \eq{3}
which has been shown to give the same accuracy from the very beginning.
However, the particular cancellation in the first loop coefficient in
$\eta_A$ at real masses $m_b$ and $m_c$ makes it unsafe the standard way of
varying the scale at which $\as$ is evaluated. From this point of view it is
more reliable to consider the case when $m_b=m_c$ and to vary the scale of
the strong coupling near $\mu=m_c$ for this quantity. This way suggests the
uncertainty of about $\pm 2\div 3\%$ in $\eta_A$  -- the number that
seems to be more reasonable as an account for the physics at the scale
$m_c$, and is supported by the numerical discussion above. The real
clarification of the value of $\eta_A$ as well as a more confident estimate
of its theoretical uncertainty at the level of a few percent is possible
only by the exact calculation of the two loop correction to $\eta_A$.

In the absence of real two loop calculations the best one has at present for
the perturbative correction factors is the one loop expression of \re{SV}
where the scale of the strong coupling is taken to be $\sqrt{m_cm_b}$:
$$
\eta_A=1+\frac{\as(\sqrt{m_bm_c})}{\pi}
\left(\frac{m_b+m_c}{m_b-m_c}\log{\frac{m_b}{m_c}}-\frac{8}{3}\right)
\simeq 0.97
$$
\beq
\eta_V=1+\frac{\as(\sqrt{m_bm_c})}{\pi}
\left(\frac{m_b+m_c}{m_b-m_c}\log{\frac{m_b}{m_c}}-\frac{2}{3}\right)
\simeq 1.02
\label{14}
\eeq
where it is put $m_c/m_b=0.27$ and $\bar{\al}_s=0.25$,
with the uncertainty of at least about $3\%$ for the axial current. This
uncertainty must be added to the possible variation in the impact of
non-perturbative corrections in the total zero recoil formfactor of $B\ra
D^*$ decay\footnote{The method of adding the various theoretical
uncertainties in quadrature adopted in \res{N} seems to be unjustified
leading to an essential overestimate of the existing theoretical accuracy.}.
\vspace*{.25cm}

4. To summarize, in the present note I discussed in more detail the arguments
that lay behind the short comment on the existed perturbative calculations
formulated previously in \res{optsr,SUV}. They suggest that a consistent LLA
or NLA in the exclusive semileptonic $b\ra c$ decays, if possible at all,
cannot produce any trustworthy impact on $\eta_{A,V}$ above a permille
level. This is noticeably below a typical size of neglected ordinary second
order corrections. Neither in such a way existing theoretical uncertainty
can be estimated. The most general arguments demonstrate that using the
strong coupling normalized at the average scale $\mu^2=m_cm_b$ in the exact
one loop expression, \eq{14}, is as good as (or even better than) the
existing next-to-leading formulae \cite{N}, and the phenomenological
application of the latter is misleading beyond the information  contained in
\eqs{14}. The only reliable way to go beyond the one loop relations is to
calculate directly two loop corrections to the above radiative
factors.

 \vspace*{.3cm}

{\bf ACKNOWLEDGMENTS:} \hspace{.4em} I am most grateful to my collaborators
I.~Bigi, M.~Shifman and A.Vainshtein for numerous discussions including the
aspects of radiative corrections, that lead in particular to the statements
made in brief previously and which are the subject of the detailed
discussion in the present note.
I acknowledge that M.~Voloshin told me about a similar symmetry arguments
made by him to M.~Neubert as early as in 1991, when discussing the proper
form of the axial formfactor of the $b\ra c$ transitions at zero recoil.

\vspace*{1.4cm}

{\large\bf Figure Caption}\vspace*{.9cm}\\
{\bf Fig.\hspace*{.1em}1\hspace*{.5em}}The second order coefficient in
$\eta_A$ as a function of the mass ratio $m_c/m_b$ as it would follow from
the ``NLA'' formulae of \res{N}, \eq{12}.  The upper curve corresponds to
$\mu=m_c$ and the lower one is for $\mu=m_b$.


\vspace{0.4cm}

\begin{thebibliography}{99}

\bibitem{optsr}
I. Bigi, M. Shifman, N.G. Uraltsev, A. Vainshtein,
{\it Preprint} CERN-TH.7250/94, May 1994.

\bibitem{SUV}
M. Shifman, N.G. Uraltsev, A. Vainshtein,
{\it Preprint} TPI-MINN-94/13-T, April 1994.

\bibitem{SV}
M. Voloshin, M. Shifman, {\it Yad. Fiz.} {\bf 47} (1988) 801
[{\it Sov. J. Nucl. Phys.} {\bf 47} (1988) 511].

\bibitem{N}
M. Neubert, {\it Phys. Rev.} {\bf D46} (1992) 2212;\\
{\it Preprint} SLAC-PUB-6263 (1993);\\
{\it Preprint} CERN-TH. 7225/94, April 1994;\\
{\it Preprint} CERN-TH.7395/94, August 1994.
\end{thebibliography}
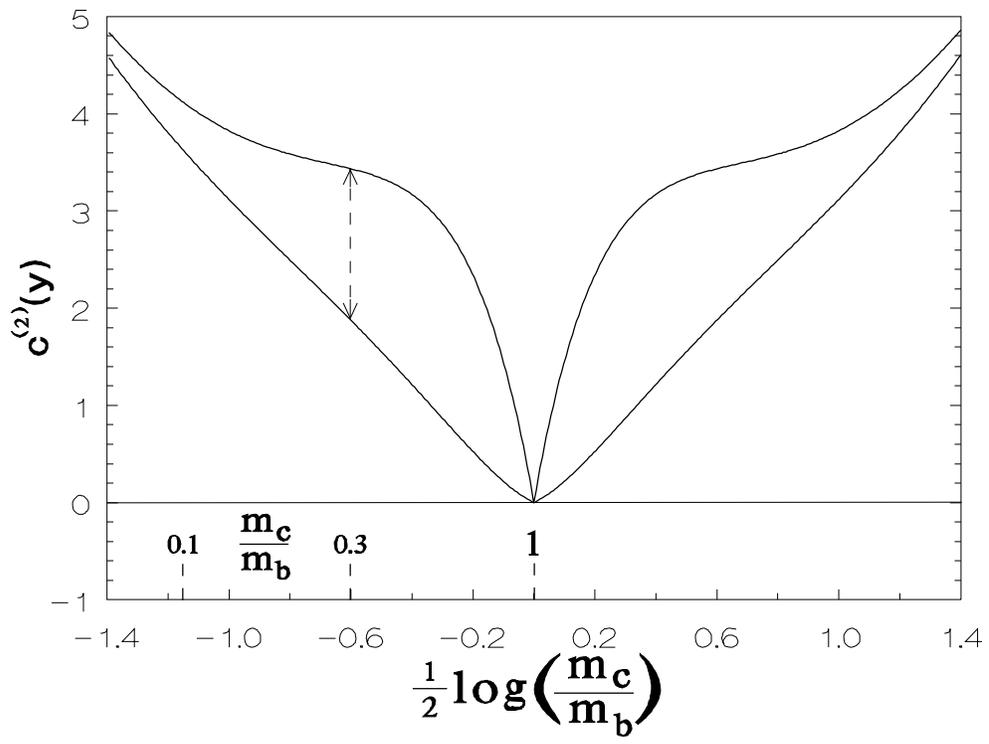
\end{document}